\def\year{2020}\relax

\documentclass[letterpaper]{article} 
\usepackage{aaai20}  
\usepackage{times}  
\usepackage{helvet} 
\usepackage{courier}  
\usepackage[hyphens]{url}  
\usepackage{graphicx} 
\usepackage{graphicx}  
\usepackage{type1cm}     
\usepackage{xspace}     
\usepackage{booktabs}     
\usepackage{multi row}     
\usepackage{bold-extra}     
\usepackage{siunitx}          
\usepackage[vlined,linesnumbered,ruled,noend]{algorithm2e}     
\usepackage{microtype}    
\usepackage{units}     
\usepackage{mathtools}     
\usepackage{amssymb}     
\usepackage{todonotes}
\usepackage[utf8]{inputenc}
\usepackage{xcolor}
\usepackage{rotating}

\newcommand{\citet}[1]{\citeauthor{#1}~\shortcite{#1}}
\newcommand{\citep}{\cite}

\newcommand{\users}{\ensuremath{\mathcal{U}}\xspace}

\newcommand{\rwc}{\ensuremath{RWC}\xspace}

\title{Falling into the Echo Chamber: the Italian Vaccination Debate on Twitter}

\author{Alessandro Cossard\thanks{After the first author, authors are in alphabetical order.}, Gianmarco De Francisci Morales, Kyriaki Kalimeri, \\ \Large \textbf{Yelena Mejova, Daniela Paolotti, Michele Starnini}\\
ISI Foundation, Turin, Italy\\
alessandro.cossard@gmail.com, \{gdfm,kyriaki.kalimeri,yelena.mejova,daniela.paolotti,michele.starnini\}@isi.it}

\begin{document}

\maketitle

\begin{abstract}
The reappearance of measles in the US and Europe, a disease considered eliminated in early 2000s, has been accompanied by a growing debate on the merits of vaccination on social media. 
In this study we examine the extent to which the vaccination debate on Twitter is conductive to potential outreach to the vaccination hesitant.
We focus on Italy, one of the countries most affected by the latest measles outbreaks. 
We discover that the vaccination skeptics, as well as the advocates, reside in their own distinct ``echo chambers''. 
The structure of these communities differs as well, with skeptics arranged in a tightly connected cluster, and advocates organizing themselves around few authoritative hubs. 
At the center of these echo chambers we find the ardent supporters, for which we build highly accurate network- and content-based classifiers (attaining 95\% cross-validated accuracy). 
Insights of this study provide several avenues for potential future interventions, including network-guided targeting, accounting for the political context, and monitoring of alternative sources of information.
\end{abstract}

\maketitle

\section{Introduction}
\label{sec:intro}

Vaccines are undoubtedly one of the most successful and cost-effective health interventions, both at individual and community level. 
Despite their effectiveness, \emph{vaccine hesitancy}, ``the reluctance or the refusal to vaccinate despite the availability of vaccination services'', potentially has the power to reverse the gains from vaccination~\cite{wiyeh2018vaccine}. 
In 2018, measles coverage has declined in 12 states of the European Union, while approximately 40\% of parents in the United States delay or refuse vaccinations for their children~\cite{smith2011parental}.
As a consequence, vaccine hesitancy has been included in the top 10 threats to global health in 2019 by the World Health Organization.\footnote{https://www.who.int/emergencies/ten-threats-to-global-health-in-2019}

In the past decade, vaccine hesitancy has been related to unwillingness to engage with scientific evidence~\cite{browne2015going}, alignment with alternative or holistic health~\cite{Kalimeri2019}, 
as well as spiritual and religious identities~\cite{kata2010postmodern}, 
anti-authoritarian worldviews~\cite{browne2015going},
conspiracy theories~\cite{jolley2014effects}, and political attitudes~\cite{yaqub2014attitudes}.
As social media -- and especially Twitter as a public platform -- becomes increasingly important in the expression of such attitudes, public health institutions are considering it as a potentially useful tool for opinion research 
and public engagement \cite{ortiz}. 
Thus, we ask, is Twitter a potentially fruitful platform for reaching out to the vaccine hesitant?

Unfortunately, with the rise of social media, an additional divisive force has come into play: the so-called \emph{echo chamber} effect, whereby users have their beliefs reinforced via interactions with like-minded peers.
This phenomenon may have roots in behavioral biases (such as selective exposure, i.e., the tendency to look for what we already agree with), and algorithm biases (which narrow the available information sources based on our digital profiles). 
As calls rise for social media-based interventions, it is crucial to ($i$) characterize the echo chambers around vaccine attitudes on social media and ($ii$) develop tools to identify those who may be targeted for potential intervention, i.e., those who are still hesitant.

In this study, we consider the case of the Italian vaccination debate, particularly interesting given that in the recent past Italy has been among the European countries with the highest number of measles cases.\footnote{https://www.ecdc.europa.eu/en/publications-data/monthly-measles-and-rubella-monitoring-report-march-2019}
This situation led to the introduction of laws imposing mandatory vaccinations for children to attend school in December 2017, which sparked an active debate.\footnote{https://www.newscientist.com/article/2196534-italy-bans-unvaccinated-children-from-schools-after-measles-outbreaks}
This debate, we find, is characterized by two clearly identifiable echo chambers, one formed by users supporting vaccination, or vaccine advocates, and the other formed by users skeptic about vaccination. 

Not only the two communities have \emph{distinct preferences of information sources} and topical subjects (with vaccine skeptics strongly favoring YouTube, similarly to their counterparts in US~\cite{monsted2019algorithmic}), but \emph{the topology of the two echo chambers differs significantly}.
In addition, the \emph{interaction between the communities is asymmetrical}, with vaccine advocates ignoring the skeptics (which may be a conscious policy decision,\footnote{https://www.wired.co.uk/article/anti-vaxxer-nhs-plan} or a missed opportunity to engage with the opposition), while being mentioned heavily by the other side, usually as a form of criticism.
The skeptic community has the typical characteristics of a vocal minority: while being smaller, its voice is disproportionately larger on Twitter.

Moreover, we find the debate to be \emph{highly politicized}, with famous politicians as the main actors, and the most predictive features of the two stances being other political attitudes, including anti-European and anti-immigration ones. 
Additionally, our network- and content-based stance classifiers work well on the ardent supporters in the center of the communities (attaining $95\%$ cross-validated accuracy), but less so for those expressing more nuanced or vague opinions at the periphery of the debate (accuracy falls to $86.7\%$), thus suggesting \emph{it may be difficult to identify those who are susceptible to changing their minds}.

\section{Related Work}
\label{sec:related}

The existence of echo chambers is a highly debated topic~\cite{dubois2018echo} 
and a rich literature addressing the siloing of political communities on social media has recently blossomed.
Echo chambers have been quantified in several controversial debates on different social media platforms~\cite{garimella2018political,del2016echo}. 
They have mostly been studied in the political context, where the sides correspond to opposing political ideologies which are nevertheless structurally similar to each other~\cite{conover2011political,garimella2018political}, 
although behavioral differences may exist~\cite{barbera2015tweeting}.
Inside these closed communities, the information spread is homogeneous and often biased, thus fostering homophilic attitudes~\cite{del2016spreading,cota2019quantifying,garimella2017effect}.
Echo chambers around the vaccination debate have recently been observed on Facebook in US~\cite{schmidt2018polarization} and in France~\cite{gargiulo2019beyond}, showing stark differences in the sources of information each side favors.
In the case of vaccination debate on Twitter in Italy, the two chambers also present a markedly different structure, as we show in this paper.

In 2012, concerns arose when groups resilient to vaccination started taking advantage of Facebook, Twitter, and YouTube to coordinate their efforts and publish content that is ``often vivid, emotionally arousing, and personal''~\cite{betsch2012opportunities}.
Since then, several studies have discovered that vaccination skeptics remain a minority, but an exceedingly vocal one, with a small fraction of users generating most of the content regarding vaccine hesitancy~\cite{monsted2019algorithmic}.
Unlike their pro-vaccination counterparts, they are more likely to link to emerging news websites and social media, and share proportionally more URLs in their tweets in general; however, picked from a more limited URL pool~\cite{chen2018interaction}. 
While vaccine advocates continue to favor traditional mainstream media sources such as newspapers and magazines~\cite{meadows2019twitter,monsted2019algorithmic}.
We find confirmatory evidence of most of these patterns in our data, not only via content analysis, but in the structure of interactions between the two groups.

Fact checking and exposure to contrarian content have been shown to be both effective~\cite{garimella2017reducing,horne2015countering} 
and counterproductive~\cite{Bail9216,nyhan2010corrections} in reducing polarization, depending on the specific setting.
Nevertheless, a first necessary step to any intervention is identifying the target population, which we also tackle in this work.
Unlike previous efforts in automatically identifying medical rumors~\cite{ghenai2017catching,ghenai2018fake}, we exploit both network and content information, and evaluate these classifiers on both users close to the center of their community and those in the periphery. 
In the past, Twitter data has proven to be useful in explaining some variation in the vaccine coverage rates, as reported by the immunization monitoring system of WHO~\cite{bello2017detecting}.
It has also been used to classify the stance of a user towards vaccination, among other controversial topics by using low-dimension projection of text-based features and unsupervised learning~\cite{stefanov2019predicting}.
Supervised learning based on deep neural networks has also been used for stance classification in the vaccine debate~\cite{monsted2019algorithmic}, achieving accuracy of 90.4\% in two-class setting.
In this work, we utilize two sources of information: firstly structure of the retweet network to perform stance classification of users within the main cluster of the conversation (guided by previous work~\cite{garimella2016quantifying}), and secondly tweet content of these users as training data to build content-based classifiers (using features inspired by~\citet{ghenai2018fake} as well as bag-of-words ones) to label users outside of this cluster.
By considering users both central to vaccination skeptic movement and on its periphery, we contribute tools to potentially identify those who may be susceptible to targeted intervention.


\section{Data collection}
\label{sec:data}

We query the Twitter Streaming API with the following Italian keywords:
\emph{vaccini, vaccinazioni, vaccinazioni obbligatorie, vaccinazioni scuola, vaccinazioni legge Lorenzin, immunodepressi scuola vaccini} (in English: vaccines, vaccinations, mandatory vaccinations, vaccinations school, vaccinations Lorenzin law, immunodeficient school vaccines), which cover the keyword ``vaccine'', as well as the school-related legislation named after the Italian former Ministry of Health, Beatrice Lorenzin.
The collection spans August 7, 2018 to April 13, 2019, encompassing 250 days, and includes \num{818668} tweets from \num{102017} users.
We use no language tag filter, thus possibly including content of languages other than Italian.
The first notable event we capture is the first stage of passing of a law in August 6, 2018 allowing unvaccinated children to attend school until 2020.\footnote{http://www.ilsole24ore.com/art/milleproroghe-primo-ok-senato-slitta-obbligo-vaccini-AEsFFgXF}
On September 10, the same law was modified allowing unvaccinated children to enter school only until March 2019.\footnote{http://www.quotidianosanita.it/governo-e-parlamento/articolo.php?articolo\_id=65251}
The rest of the volume is generated around the following notable articles, government communications, and school deadlines.

To check for the presence of bots, we examine the top 20 users by posting frequency, and we find that all of them are on the skeptical side.
Upon manual inspection, we find that most accounts do not behave like bots, as they show consistent ideological persuasion and produce original content.
Conversely, out of the 10 most retweeted tweets, we find only one which is skeptical.
Upon manual checking, we conclude that none of these tweets was produced by bots.
Therefore, we choose not to exclude any account from the analysis, so as to maximally preserve the structure of the conversation.

In addition, we use the Twitter User Timeline API to query for each user's historical data. 
After filtering and user selection 
we proceed to query for the timelines of \num{7152} users in the giant connected component (GCC) of the retweet network, thus resulting in \num{43317280} tweets. 
We then create a similar dataset for a sample of 300 non-GCC users, thus resulting in \num{257082} tweets.
We dub this dataset \emph{historical data}.
Finally, we use the Twitter Followers API to collect the lists of users who follow or are followed by any user in our dataset, thus collecting an average of \num{5538} neighbors per user.

The data used in this research was collected via the public Twitter API for streaming and user history collection, and thus complies with the privacy restrictions of the platform, which excludes accounts that are restricted or deleted. 
The dataset is available upon request, as per Terms of Service of Twitter.

\textbf{Network construction.} The \emph{retweet network} is a weighted directed network where nodes represent users and the weight of an edge from node $u$ to node $v$ represents the number of times that user $u$ retweets user $v$. 
We exclude all edges with a weight smaller than two to reduce the noise in the data~\cite{garimella2018quantifying}.
The resulting network has \num{19193} users and \num{69888} edges.
The retweet network is our main tool to understand the stance of the users, as described in the Network-based user classification Section.
In particular, we focus on the GCC of this network, which includes more than $95\%$ of the users (denoted with \users). 

The \emph{mention network} is a weighted directed network in which nodes represent users and edges represent mentions, i.e. the action of including a username in a tweet (together with \textit{replies}).
The weight of an edge from node $u$ to node $v$ represents the number of times user $u$ mentions user $v$. 
We do not apply any threshold to the edge weight. 
The resulting network has \num{49488} users and \num{242858} edges.
If we focus on users in the GCC of the retweet network \users, it reduces to \num{13573} nodes and \num{115873} edges. 

Finally, the \emph{follow network} is a directed network in which nodes represent users and an edge from node $u$ to node $v$ represents user $u$ following user $v$. 
Given the rate limitations of the Twitter API, we reconstruct the follower network only for users in \users, resulting in \num{17650} users and \num{2195499} edges. 

The main properties of the three networks are showed in Table~\ref{tab:netstats}. 
One can see that the networks have similar size, but the follower network is much more dense and reciprocal than the other two. 
Note also that the mention network shows a lower Random Walk Controversy score~\cite{garimella2018quantifying}, indicating that it presents more connections across the two sides (see the Topology of the Echo-chambers Section).

\begin{table}[t]
\centering
\caption{Statistics of the retweet, mention, and follower networks: number of users $N$, number of classified users $N_C$, number of edges $E$, and reciprocity $\rho$, and Random Walk Controversy score \rwc}
\label{tab:netstats}
\resizebox{\columnwidth}{!}{
\begin{tabular}{lrrrrr}
  \toprule
  \textbf{Network} & \textbf{$N$} & \textbf{$N_C$} & \textbf{$E$}  & \textbf{$\rho$} & \textbf{\rwc} \\ 
  \midrule
  Retweet & \num{19193} & \num{18363} & \num{69888}  & \num{0.042} & \num{0.803} \\
  Mention & \num{49488} & \num{13573} & \num{242858}  & \num{0.081} & \num{0.336} \\ 
  Follower & \num{17650} & \num{17650} & \num{2195499}  & \num{0.355} & \num{0.704} \\
  \bottomrule
\end{tabular}
}
\end{table}


\textbf{Network-based user classification.} Retweets (the action of propagating the original message without modification) are often understood as endorsement of the opinion expressed in the retweeted message~\cite{garimella2016quantifying}.
As such, they can be used to identify groups of Twitter users which take a particular side of a controversial debate.
Following a recent work on quantifying controversy in online speech~\cite{garimella2016quantifying}, we use the structure of the retweet network to identify the leaning of users in the vaccination debate. 

First, we note that the GCC of the retweet network is roughly organized into two large, clearly identifiable groups of users, see Figure 3 (left). 
Furthermore, we test this separation by running several community detection algorithms, which consistently identify two large communities. 
Therefore, we quantify this separation by applying a graph partitioning algorithm, which comprises users in \users. 
Following previous works, we use METIS~\cite{karypis1998fast}, a multilevel graph partitioning algorithm.
We bi-partition the graph repeatedly $100$ times with different random seeds to get an ensemble of partition assignments for each node.
We then use the average partition assignment for each node across the repetitions as a leaning score of the corresponding user.
Thus, the leaning score $x_{u} \in [0,1]$ represents how likely user $u \in \users$ is to appear in one of the two partitions, \num{0} or \num{1}, the algorithm finds.
Assuming a Bernoulli distribution of the assignments, we tune the hyper-parameter of the algorithm (the relative size of the partitions) to optimize the number of users with a score within a $95\%$ confidence interval from either extreme ($0$ and $1$).
We find the optimal proportion of two sides to be $1.00$:$1.54$ (that is, one side is $54\%$ larger than the other).

By following this methodology, we obtain scores for $N_C=\num{18363}$ users, such that each user is characterized by their leaning with respect to vaccination.
As the output of the partitioning algorithm lacks semantics, we examine a selection of most prominent users in each partition and assign their stances to all users of that partition.
This approach results in \num{6065} users classified as as vaccine skeptic (having leaning score $\approx 0$), and \num{11334} as vaccine advocates (having leaning score $\approx 1$).
The fact that $94.7\%$ of users are assigned into one of the two extremes suggests a highly divided network (as we illustrate further in the paper).
However, we also examine the remaining $5.3\%$ of the users (in blue in the figure) who have not been consistently assigned to either side.
They discuss vaccination as it pertains to pets -- cats and dogs -- and they do not display a strong opinion concerning human vaccination.

The reader may argue about the propagation of the leaning from a handful of prominent users to the whole partition.
To verify the accuracy of our method, we manually evaluate a sample of the users, by randomly selecting \num{100} users and having \num{5} annotators proficient in Italian and familiar with the vaccination debate label them.
We compare the labels among pairs of annotators using Cohen's Kappa, and find a high level of agreement between the annotators ($\kappa=0.91$).
Comparing the discretized leanings to the manual labels, the partitioning algorithm achieves $95.74$ accuracy ($95\%$ CI of $\left[88.7, 98.4\right]\%$), which is a testament to the power of network-driven controversy analysis.
Thus, we use the labels determined from the partitioning of the retweet network in the following analysis of the vaccination discussion (and to color the other networks of Figure \ref{fig:networkplot}).

\begin{figure}[tbp]
\centering
\includegraphics[width=0.90\linewidth]{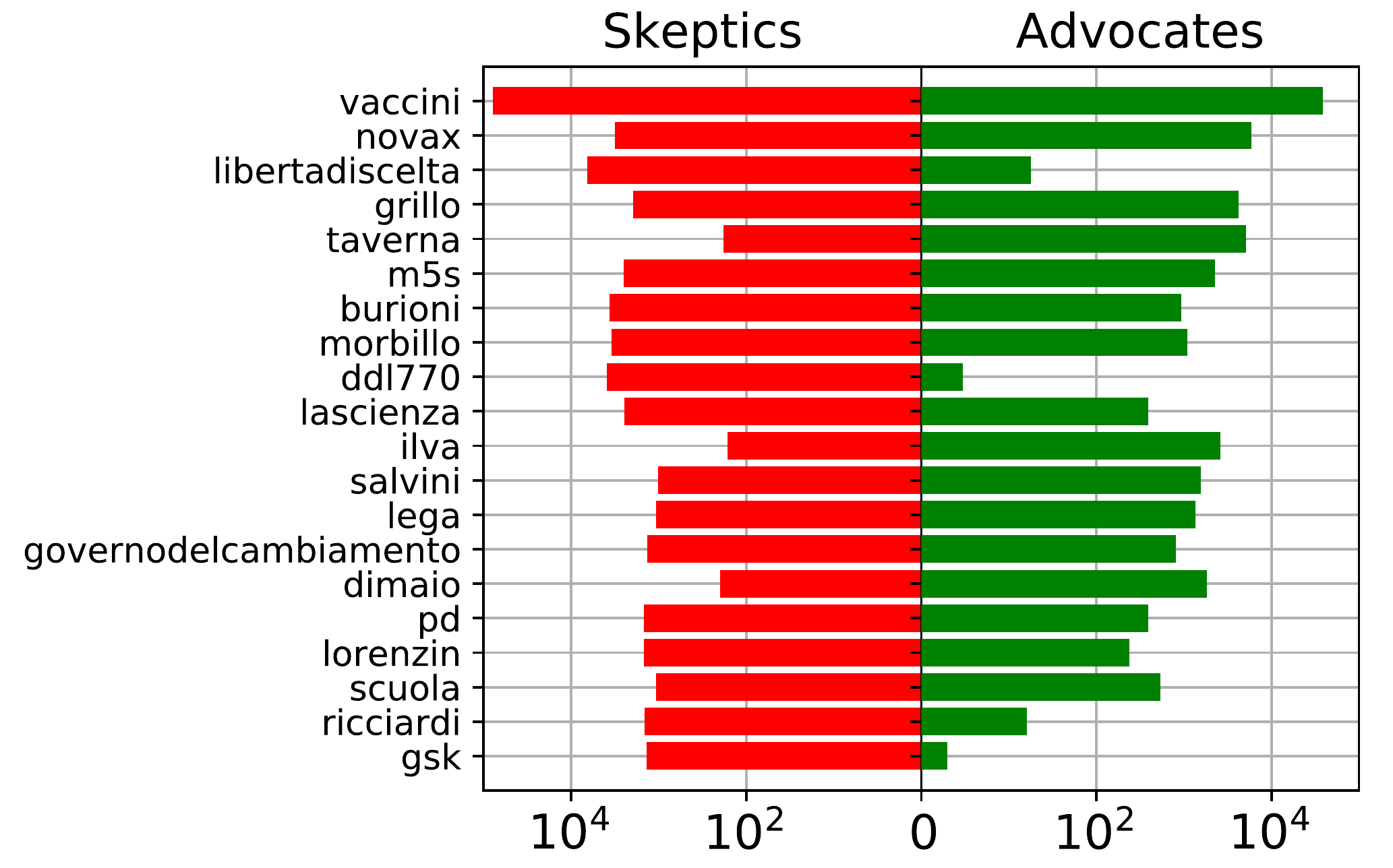}
\caption{Frequency of hashtags, ranked by total usage.}
\label{fig:freqhashtags}
\end{figure}

\section{Debate Content \& Context}
\label{sec:debate}

We now proceed to examine the content of their arguments, as seen through hashtags (as semantically cohesive tokens) and URLs they share. 
Interestingly, the rate of posting for the two sides is markedly different, with \num{227669} tweets posted by advocates, \num{20.1} tweets per user on average, and \num{368065} posted by the smaller group of skeptics, \num{60.7} tweets per user, thus signaling a behavioral difference between the two sides. 

\textbf{Text \& Hashtags.} Hashtags have been widely used as semantically cohesive tokens within tweets that warrant special attention. 
Figure~\ref{fig:freqhashtags} shows the difference in usage of the most popular hashtags by the two sides. 
Note that the two groups are rather different in size (the advocates group is 54\% larger than the skeptics), and yet the amount of hashtags used is comparable. This result indicates a higher hashtag usage by the skeptics, further evidence that the skeptics side is notably more vocal than the advocates side.
Interestingly, \#novax is used by both sides, which shows that both sides are aware of the vaccination skeptic side. 
However, \#libertadiscelta (``freedom of choice'') is used almost exclusively by the skeptics, with advocates not addressing this topic at all. 
Another popular hashtag on the skeptic side is \#ddl770, a decree in amendment of existing law which required children to be mandatorily vaccinated to allow school attendance.
The amendment reduces this requirement to a self-declaration of the parents about the compliance with vaccination requirements. 
Conversely, \#taverna, used mostly by advocates, refers to Paola Taverna (@PaolaTavernaM5S), politician of the `Five Star Movement' (M5S) and vice-president of the senate, who declared that ``when she was young she would get immunity [to measles] by visiting an ill cousin''.\footnote{http://www.ilgiornale.it/news/cronache/vaccini-quando-taverna-diceva-piccola-mi-immunizzavo-1562998.html}

\begin{figure}[tbp]
\centering
\includegraphics[width=0.90\linewidth]{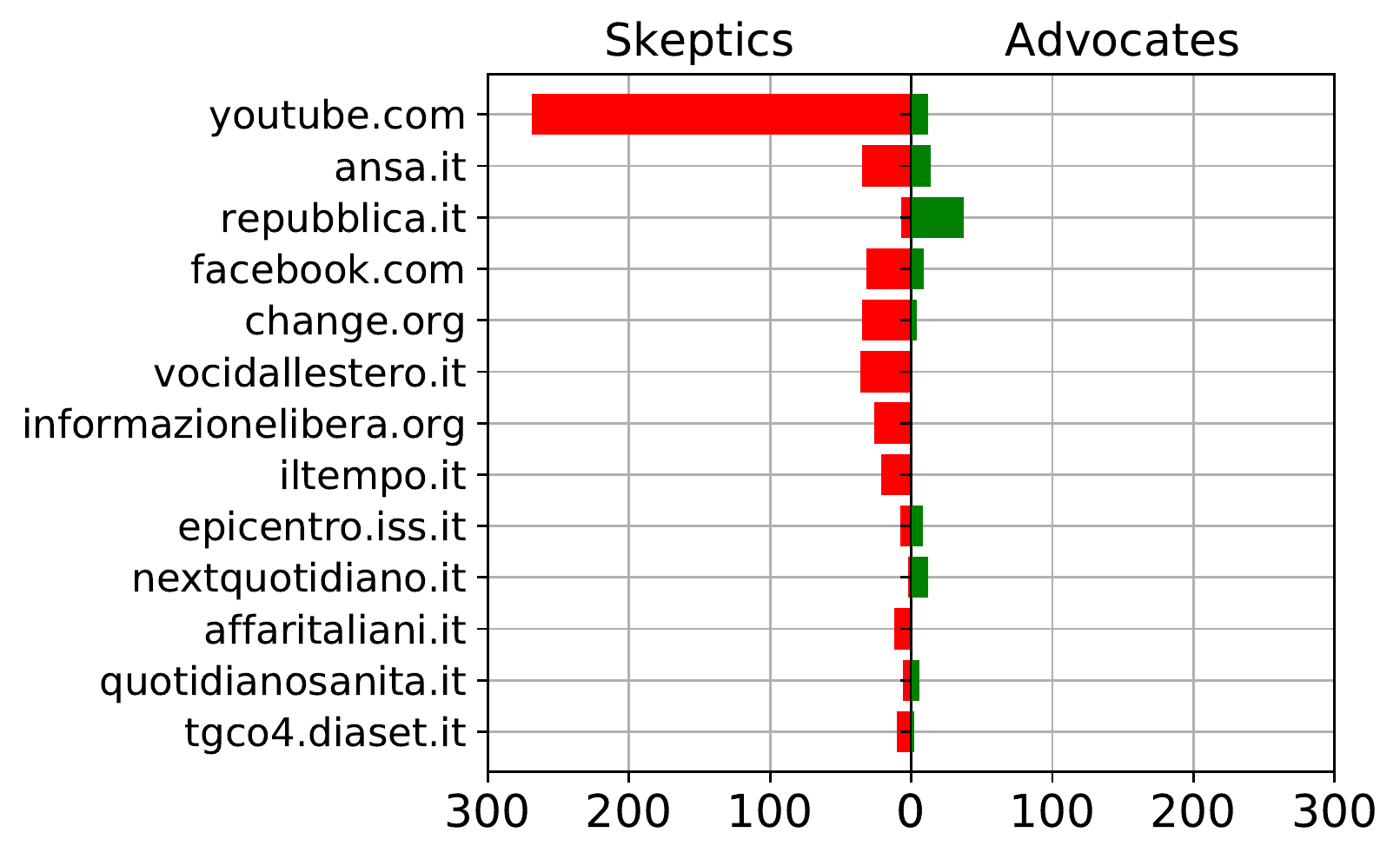}
\caption{Frequency of URL domains, ranked by total usage.}
\label{fig:frequrls}
\end{figure}

\textbf{URLs.} The source of a shared piece of information is an important signal about its quality. 
$14.1\%$ of all tweets contain a URL, with a vast majority of these being shortened, mostly by Twitter itself, and pointing to content within Twitter. 
The use of URLs is uneven between the two sides of the debate, with \num{975} unique URLs shared by the skeptics, compared to only \num{353} on the advocate side, which underscores again the vocal nature of skeptic side.
We expand the URLs which do not point back to Twitter's content. 
The most frequent domains of these URLs, ranked by total usage, are shown in Figure~\ref{fig:frequrls}.
The horizontal axis represents the number of tweets which contain a URL from a given domain on each side.

The domain most shared by skeptics is YouTube, which has recently been shown to provide ample anti-vaccination content in Italy~\cite{donzelli2018misinformation,covolo2017arguments}. The three most shared videos by skeptics are interviews to a skeptic physician, Stefano Montanari, in which he denounces the dangers of vaccines.  
Facebook is another popular social media resource for the skeptic side.
The Facebook URLs shared by skeptics point mainly to Facebook groups against vaccines.
These findings are consistent with a recent study of the vaccination debate in United States~\cite{monsted2019algorithmic}, which finds that these resources are exceedingly popular with skeptics.
As main information source we find vocidallestero.it, an alternative blog platform, with a strongly euro-skeptic, anti-globalist orientation.
The most shared domain by advocates is repubblica.it, one of the most popular general-interest newspaper in Italy, especially in its online version.
These differences in the preferences of media outlets among the two groups confirm that skeptics mostly rely on and disseminate unofficial (and often unverified) sources of information, while advocates prefer to share journalistic articles from well known and recognized media outlets.
Interestingly, among the URLs shared by skeptics we also find epicentro.iss.it which is the official Web page of the national institute of public health.
Links to the official vaccination rates for measles are in fact used by skeptics to support their claim that vaccination rates are not decreasing.


\begin{figure*}[tbp]
\centering
\includegraphics[width=0.33\linewidth]{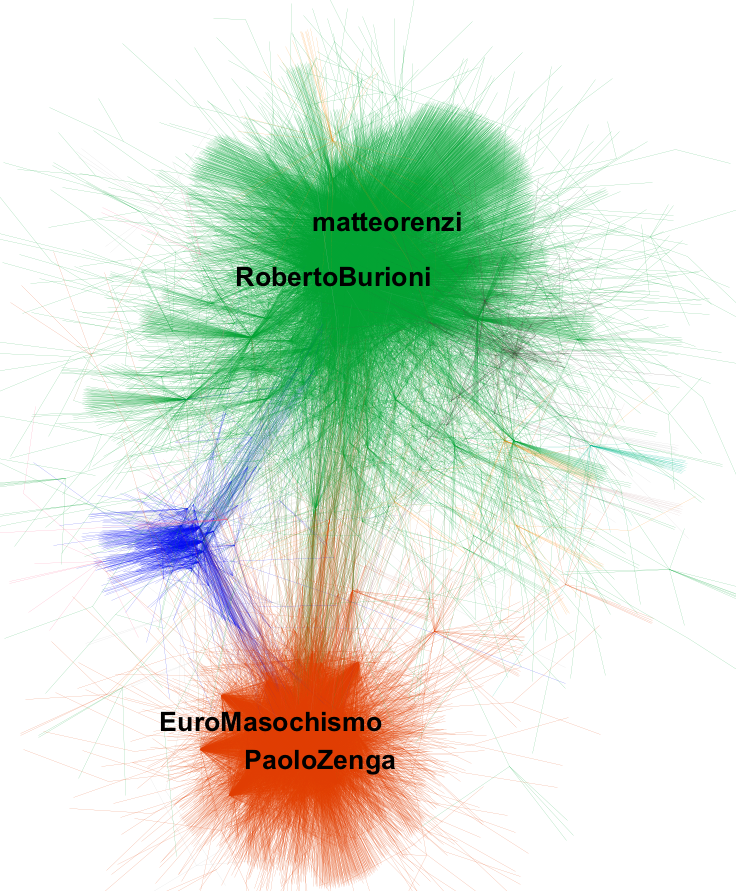}
\includegraphics[width=0.33\linewidth]{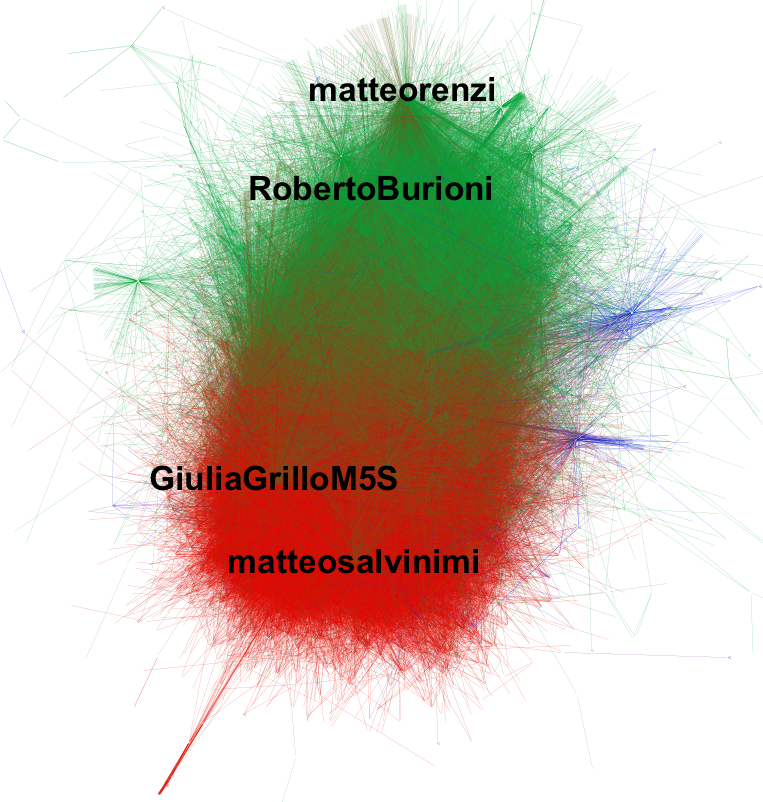}
\includegraphics[width=0.33\linewidth]{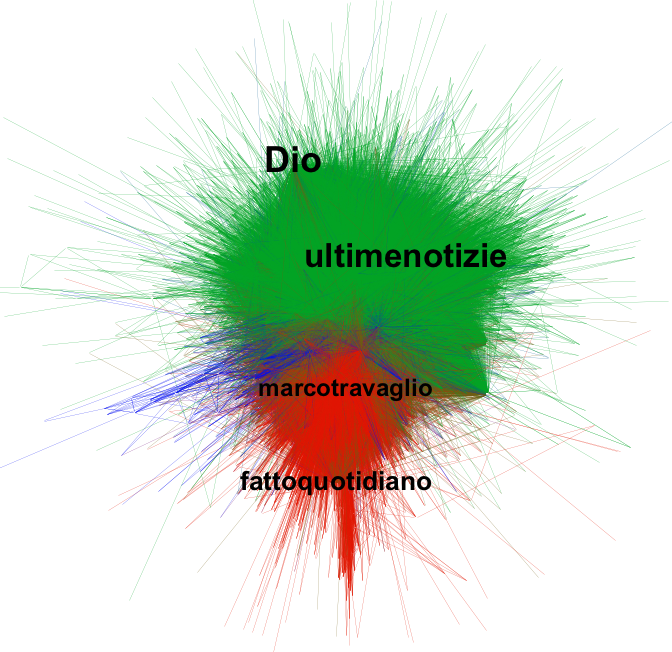}
\caption{Giant connected component of the retweet (left), mention (center), and follow (right) networks, colored by the leaning score of retweet network partitioning (green: advocates, red: skeptics, blue: pet owners, best seen in color).} 
\label{fig:networkplot}
\end{figure*}

\begin{figure*}[tb]
\centering
\includegraphics[width=0.31\linewidth]{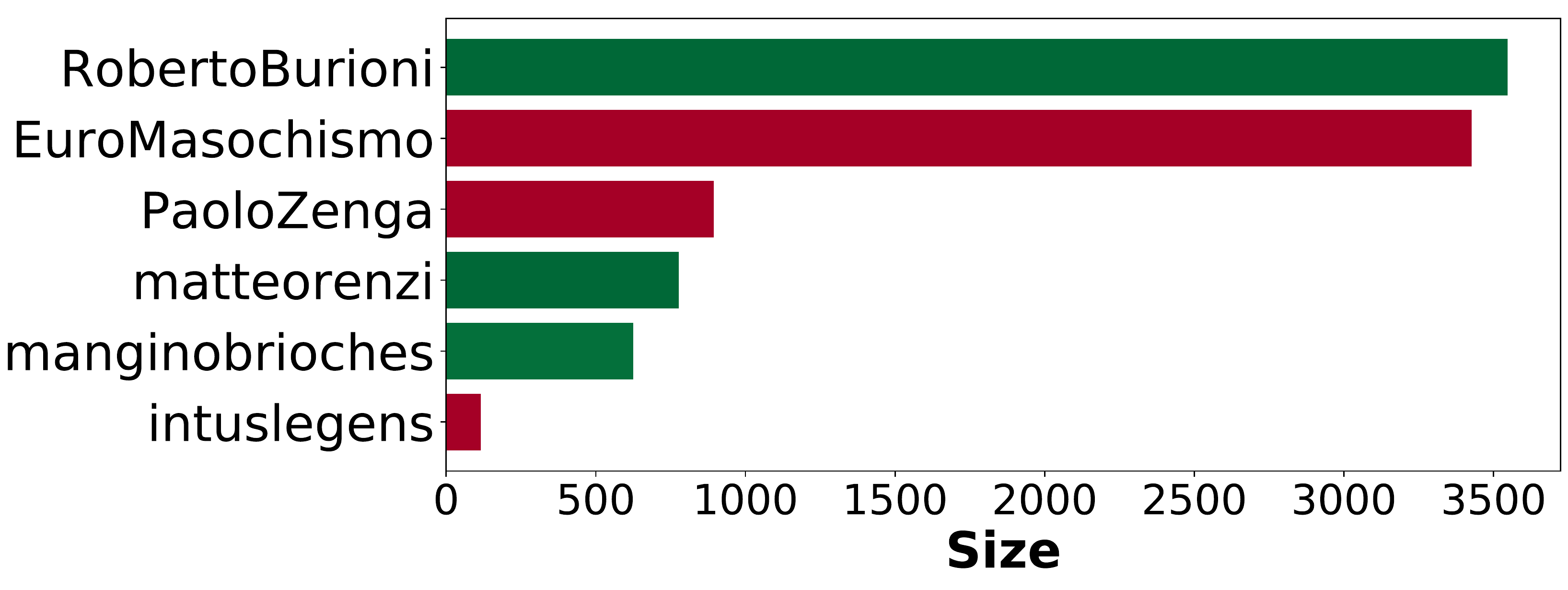}
\includegraphics[width=0.31\linewidth]{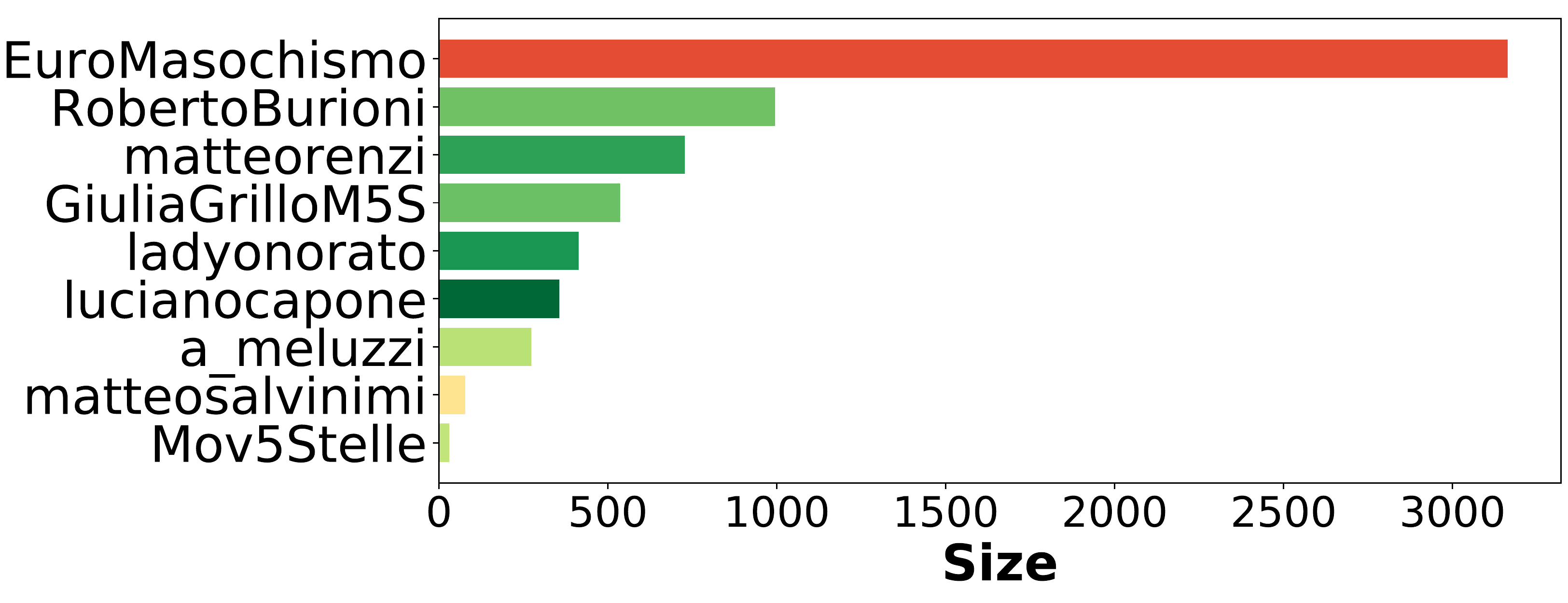}
\includegraphics[width=0.32\linewidth]{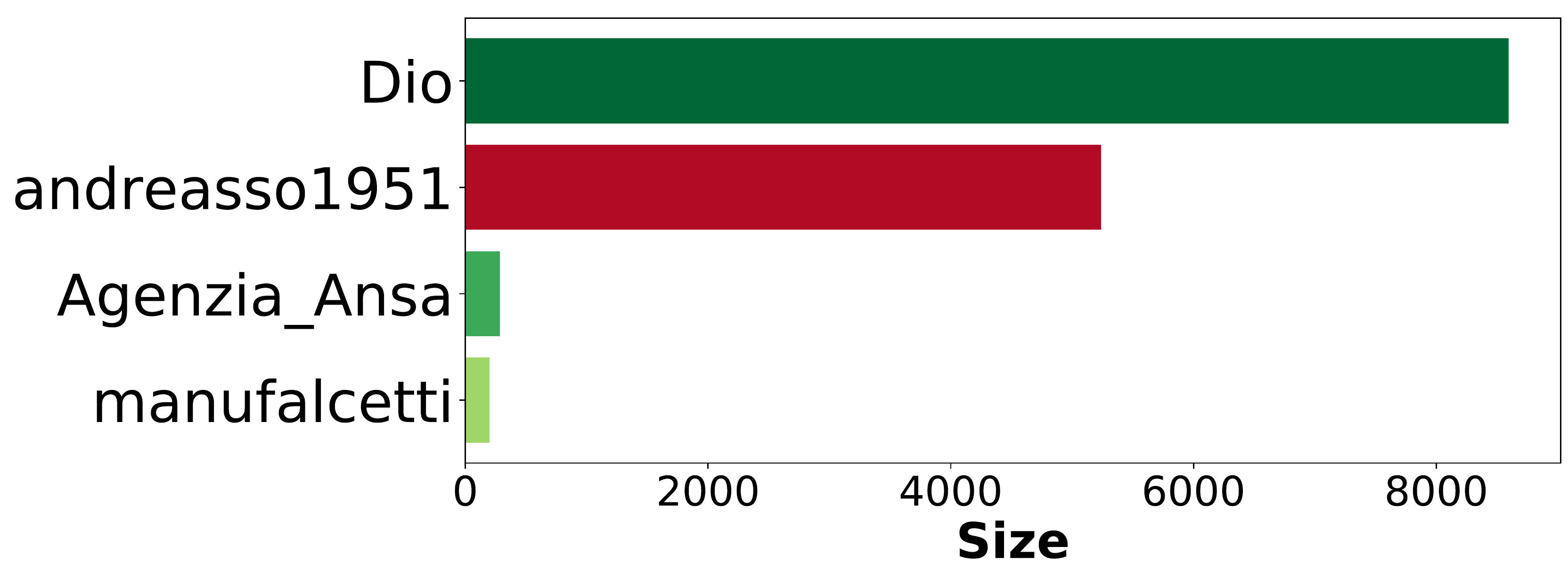}
\includegraphics[width=0.0250\linewidth]{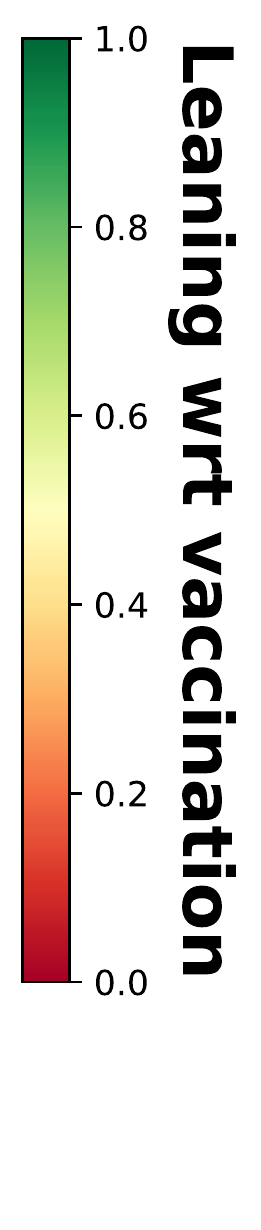}
\caption{Size and average leaning (green for advocates, red for skeptics) of Infomap communities for retweet (left), mention (center), and follow (right) networks (best seen in color).}
\label{fig:communities}
\end{figure*}

\section{Topology of the echo chambers}
\label{sec:inseide-EC}

Whereas the previous section shows the distinct content characteristics of the two sides of vaccination debate, in this section we ask to what extent do the two sides interact by following, retweeting, or mentioning each other. 
That is, does Twitter foster an actual dialogue between users with differing opinions, or is there evidence of the interaction segregation characteristic of echo chambers?

Figure~\ref{fig:networkplot} shows the retweet, mention, and follow networks, constructed as explained in the Network Construction Section. 
Each node is colored according to the user's leaning with respect to the vaccination debate, inferred through the classifier based on the topology of the retweet network: green for vaccination advocates, red for skeptics, and blue for those in-between. 
For the mention and follow networks, we show only the subgraph induced by classified users \users.
Visually, we can observe that the retweet network is clearly split into two communities of different color, as expected by the definition of the classifier. 
In this section we quantify the presence of echo chambers in each network, discuss their topological features, and show striking difference between advocate and skeptic communities.


\textbf{Community structure.} We examine the structure of the networks by applying Infomap \cite{rosvall2008maps}, a well-known community detection algorithm for directed networks. 
As a sensitivity check, we test other community detection algorithms such as the one proposed by Louvain  \cite{blondel2008fast} and the one proposed by Clauset-Newman-Moore \cite{clauset2004finding} with similar results.
Figure \ref{fig:communities} shows the largest communities for each network, colored by the average leaning of its members, and labeled by the user with the largest in-degree.
While communities characterized by very strong average leaning, indicated by dark green and red coloration of the bars, are expected in the retweet network by the definition of the classifier, they are not obvious in the other networks. 

On the advocates side, the most retweeted users are Roberto Burioni  (@RobertoBurioni), a physician, university teacher, and pro-vaccination activist, and Matteo Renzi (@MatteoRenzi), the former prime minister of Italy.
On the skeptics side we find Paolo Zenga (@PaoloZenga) and @EuroMasochismo.
We observe the lack of credentials and increased anonymity of the users popular in the skeptic communities, which in addition to anti-vaccination content also sport anti-European and nationalist themes, together with conspiracy theories, and far-right messages.

Mention network communities center around similar users, however the structure of the two sides differs substantially, with vaccine skeptics loosely grouped into a single community (note the lighter red, indicating the community sometimes mentions users with different leanings), whereas vaccine advocates are broken down into a handful of communities.
The fractured nature of the way vaccine advocates mention each other shows the lack of a coordinated effort or a message leader (although Roberto Burioni is potential contender for such role).
In contrast, the fact that vaccination skeptics mention each other so intensely points to a well-connected and self-aware community, at the center of which are prominent politicians. 
such as 
the former minister of health Giulia Grillo from the M5S, a party which has been associated with skeptical positions around vaccines,\footnote{http://www.nytimes.com/2017/05/02/opinion/vaccination-populism-politics-and-measles.html}
and the former minister of the interior Matteo Salvini from the `Lega Nord' party who famously declared that ``10 mandatory vaccines are useless and in many cases dangerous if not outright harmful''.\footnote{http://www.ilsole24ore.com/art/salvini-contro-vaccini-dieci-sono-inutili-e-dannosi-poi-ringrazia-due-paladini-no-vax-AET4xpAF}

Finally, we examine the structure of the following relationship, which represents the sources of information the users interested in vaccinations follow. 
Infomap cleanly separates the two sides into two large communities, with the rest being of negligible size.
At the center of vaccine advocates, we find @Dio (god in Italian), a satirical account with more than 800k followers, and @ultimenotizie, a breaking news aggregator with 98k followers.
On the skeptics side, Marco Travaglio (@marcotravaglio) an Italian journalist, and @fattoquotidiano a newspaper whose director is the same Marco Travaglio.
The newspaper has been criticized for a heavy pro-M5S bias, and for publishing inaccurate information.\footnote{https://www.repubblica.it/politica/2018/08/10/news/il\_metodo\_ travaglio\_dieci\_anni\_dopo-203789150/}
It has also hosted opinion pieces that link measles vaccine and autism.\footnote{https://www.ilfattoquotidiano.it/2014/07/11/vaccini-pediatrici-e-autismo-le-ricerche-di-singh/1057854}
Note that neither of these accounts appear frequently when we examine the retweet and mention relationships.
Still, the fact that the accounts the two sides of the vaccination debate choose to follow are so disjoint provides another view of the echo chamber in which they reside.

\textbf{Quantifying controversy of the networks.}
We use the Random Walk Controversy (\rwc) score introduced by~\citet{garimella2016quantifying}.
\rwc is a measure of how controversial a topic discussed on social media is, i.e., how polarized a discussion it creates among the users.
It is a network-based measure which relies on an endorsement network, a network where a link from $u$ to $v$ implies that $u$ endorses the opinion of $v$ on the given topic.
\rwc is a number in $[0,1]$ which represents the difference in probability for an average user in the network to be exposed to information from their own side versus from the opposing side.
As such, an \rwc close to $1$ represents a controversial topic with two distinct groups that do not endorse each other's opinion, while an \rwc close to $0$ represents a non-controversial topic where both opinions are equally likely to be received. 
The formula is $\rwc = P_{XX}P_{YY} - P_{XY}P_{YX}$, where $X$ and $Y$ are the two sides, and $P_{AB}$ is the probability of a random walker starting from $A$ to reach $B$.

We compute \rwc for all three networks, results are reported in Table \ref{tab:netstats}.
Both the retweet network and the topic-induced follow networks are examples of an endorsement network on Twitter, as used in the original work.
As expected, the retweet and follow networks show a highly controversial topic, as indicated by the values of the \rwc score, while the mention network shows a lower score, as it presents more connections across the two sides.
The mention network, however, does not necessarily encode endorsement.
Indeed, \citet{conover2011political} show that mentions are often used to speak to (and attack) the opposing side.
By looking at the components of the score for the mentions, the probability of skeptic information to reach the advocate side $P_{SA}$ (0.46) is much higher than the obverse one $P_{AS}$ (0.18). 
That is, skeptics have a higher tendency to mention advocates, but this attention is not reciprocated, an asymmetry that we shall find again in the next section.

\begin{figure}[tbp]
\centering
\includegraphics[width=\columnwidth]{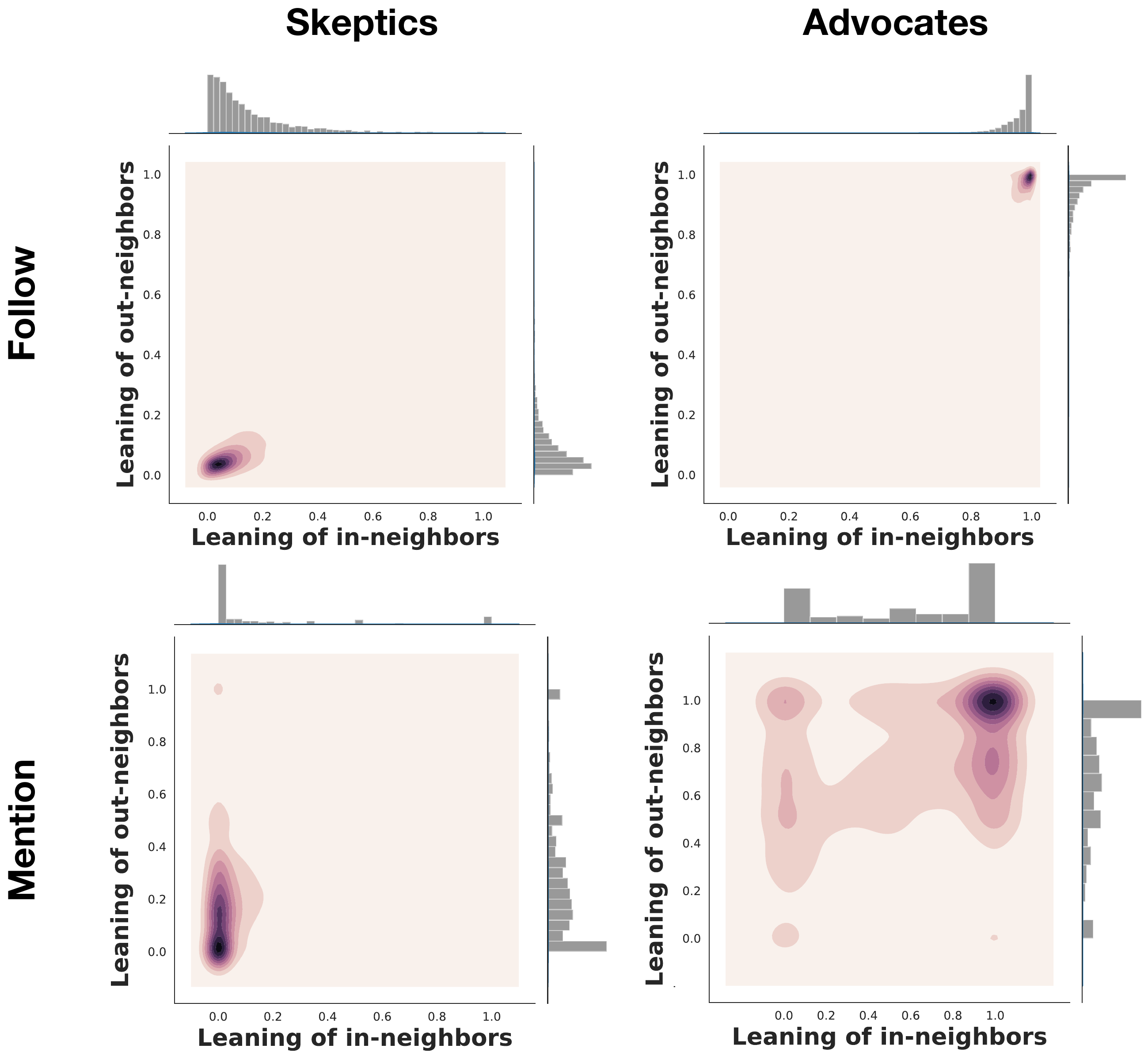}
\caption{Joint distribution of the average leaning of in-neighbors and out-neighbors.}
\label{fig:galaxies}
\end{figure}

\begin{figure}[tbp]
\centering
\includegraphics[width=\columnwidth]{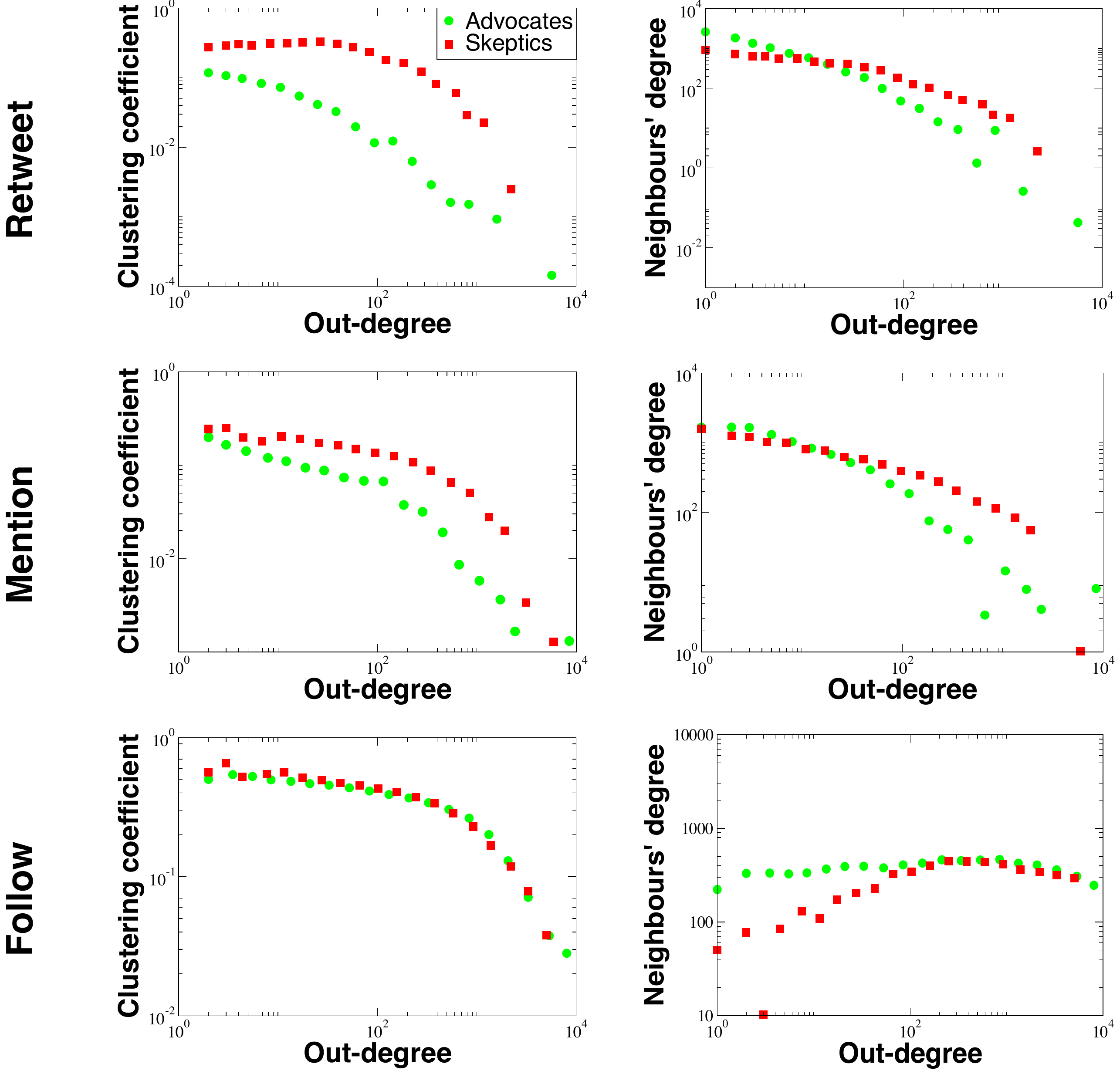}
\caption{Degree Assortativity and Clustering Spectrum as a function of the out-degree. Plots for the in-degree are qualitatively similar (omitted due to space constraints).}
\label{fig:topprop}
\end{figure}

\textbf{Asymmetry of the two chambers.} We have shown that the largest communities that emerge in the three networks are characterized by a marked average leaning towards one of the sides, and that the networks are strongly polarized.
These results are hints of the presence of echo chambers: groups of like-minded users who are more likely to interact within the group.
However, echo chambers can be more precisely quantified by relating the leaning score $x_i$ of a user $i$  with the average leaning of their neighborhood \cite{cota2019quantifying}, defined as $\langle x_i\rangle^{NN} \equiv \frac{1}{k_i} \sum_j a_{ij} x_j$, where $a_{ij}$ represents the adjacency matrix of the interaction network and $k_i \equiv \sum_j a_{ij}$ defines the degree of node $i$. 
Since we consider directed networks, we distinguish between the average leaning of in-neighbors and out-neighbors of each node. 

Figure~\ref{fig:galaxies} shows a density scatter plot between the average leaning of in-neighbors and the average leaning of out-neighbors, separately for vaccines advocates and skeptics, for both mention and follow networks.\footnote{We omit showing the result for the retweet network as we use it to compute the leaning score, and therefore to avoid an obvious result due to circular reasoning.}
All plots show color-coded contour maps, which represent the number of users with a given combination of leanings of their in- and out-neighborhoods: the darker the area in the map, the larger the density of users in that area.
Both advocate and skeptic users follow and are followed by users sharing very similar leaning (top row of Figure~\ref{fig:galaxies}), therefore confirming the echo chamber phenomenon: users receive information (via the follow relation) mostly by peers who share their leaning. 
A slightly different behavior is observed in the mention network (bottom row of Figure~\ref{fig:galaxies}).
While users are still more likely to mention and be mentioned by others with similar leaning, striking differences emerge between the two sides: vaccine advocates are more likely to interact between themselves, while vaccine skeptics frequently mention users from the other side.
Upon manual examination, we find the tweets exchanged across the sides are mostly used to convey attacks and adversary messages, such as portraying the opposing side in a bad light.

The difference between skeptic and advocate communities can be further analyzed by looking at the structure of the two echo chambers. 
Figure~\ref{fig:topprop} shows the degree assortativity and clustering spectrum of retweet, follow, and mention networks, for both advocates and skeptics. 
For all networks under consideration, the clustering coefficient strongly decreases with the out-degree (left column of Figure~\ref{fig:topprop}), which indicates that more active users have a smaller clustering coefficient.
Nevertheless, for mention and retweet networks, the clustering coefficient of skeptics is much larger than the one of advocates, especially for highly connected nodes (hubs).
A similar behavior is observed for in-degree (omitted due to space constraints), which represents user popularity.
These results indicate that the vaccine skeptic community is more tightly connected. 

Looking now at the right column of Figure~\ref{fig:topprop}, for retweet and mention networks, the average degree of nearest neighbors is a decreasing function of the out-degree, which follows a power-like function form, thus indicating that these networks are both disassortative. 
Again, such behavior differs between the two communities: advocates show a stronger dissasortative character than skeptics.
Disassortativity indicates that advocate hubs are likely to be connected to low-degree nodes, in a star-like structure. 

By taking into account these results, we surmise that the topology of the two echo chambers is significantly different: the skeptic community is more tightly connected, with a mesh network topology, while vaccine advocates show a hierarchical structure around hubs, such as Matteo Renzi 
and Roberto Burioni. 

\section{Falling into the echo chamber: Prediction task}
\label{sec:classification}

The analyses described thus far lack a temporal dimension, as they look at a snapshot of the conversation. 
For the purposes of a potential intervention, however, it is important to find the Twitter users who may be vulnerable to ``falling into the echo chamber'', and especially joining the vaccination skeptic conversation. 
Therefore, we ask whether it is possible to predict the side a given user will join (operationalized as retweeting a user of known leaning), based on the history of their tweets in the historical data.

We cast this task as a standard binary classification one.
The set of examples is given by the users \users (those in the GCC of the retweet network).
The label is given by the leaning of the user ($\{0,1\}$ for skeptic and advocate, respectively) determined as per Network-based User Classification Section.
However, we restrict to users who retweet only users on a single side (i.e., we exclude users who retweet both sides).
Given the history of tweets of a user $u \in \users$,
we define the \emph{split point} as the time $t$ when $u$ first retweets another user $v \in \users$ such that $v$ has a known leaning, and take only users with split points no earlier than 2017.
Then, we define the \emph{history} of a user $u$ as all the tweets in the historical data posted before the split point $t$.
We use the history of a user to extract features for the prediction task as detailed in the next section.
Out of all the labeled users, we include \num{4813} vaccine advocates and \num{2066} vaccine skeptics in this experiment (for a majority baseline accuracy of \num{0.68}).


\textbf{Features.} We begin by crafting features inspired by previous work on misinformation~\cite{ghenai2018fake}, thus obtaining the following $16$ features (normalized where appropriate by the number of tweets) we dub \emph{aggregate}, computed for each user:

\begin{itemize}
  \item  User (5): account age in days, total number of tweets, tweet rate, number of followers, number of friends;
  \item  Twitter specific (4): proportion of tweets that are retweets, have mentions, have hashtags, or have a URL;
  \item  Lexical (7): number of character, upper case characters, verbs, nouns, articles, question marks, exclamation marks.
\end{itemize}

All the features are statistically significantly different between the two sides, according to both Student's t-test and Mann-Whitney U test, at a significance level of $p < 0.001$.

In addition, we create bag-of-words (\emph{BoW}) features by treating all the tweets by a given user as a document.
We preprocess the tweets by removing URLs, mentions, and stopwords, lemmatize the words to retain only nouns, adjectives, verbs, and adverbs, and apply $10$ as the frequency cut-off for the resulting vocabulary, thus yielding 73k features, on which we apply TF-IDF weighting.

\textbf{Performance Results.} For the prediction model, we use off-the-shelf classifiers as provided by scikit-learn,\footnote{http://scikit-learn.org} Random Forest and Logistic Regression (with L2 regularization). 
We use 5-fold cross-validation to compute accuracy, AUC (area under the ROC curve), and F1 score (harmonic mean of precision and recall), shown in Table \ref{tab:classresults}. 
The aggregate features are best used by Random Forest (whereas Logistic Regression performs close to random baseline). 
The BoW features, instead, are best used by the Logistic Regression model. 
The hyper-parameters for Random Forest are tuned via cross-validated grid search.
The resulting classifier uses balanced class weights, $200$ trees, a minimum of 8 examples per split, and a maximum tree depth of 30. 

\begin{table}[t]
\centering
\caption{Prediction performance of Random Forest (RF) and Logistic Regression (LR) models when using aggregate and BoW features. Trained on users classified via retweet network, tested via 5-fold cross-validation.}
\label{tab:classresults}
\begin{tabular}{l ccc}
\toprule
 & \textbf{Accuracy} & \textbf{AUC} & \textbf{F1} \\
\midrule
RF, aggregate  & 0.854 & 0.806 & 0.898 \\
LR, aggregate  & 0.672 & 0.626 & 0.758 \\
RF, BoW & 0.908 & 0.959 & 0.937 \\
LR, BoW & 0.954 & 0.975 & 0.967 \\
\bottomrule
\end{tabular}
\end{table}

We find that the larger BoW feature set performs the best, whereas the hand-crafted aggregate features are only useful when using Random Forest. 
In fact, these $16$ features show a marked improvement over the majority baseline of \num{0.68} accuracy, which means that there is a strong signal in simple lexical features such as grammar, writing style, and tweeting behavior.
We also combine the two feature vectors, and find the performance increase marginal beyond what achieved by the BoW feature set.

The prediction performance achieved by the Logistic Regression model with BoW features is very high.
However, it relies on the presence of specific words and topics.
Therefore, concept drift may deteriorate performance over time.
Instead, the aggregate features, while not able to achieve the same level of accuracy, are more portable and robust, as they rely on stylistic and behavioral patterns.


\textbf{Feature Importance.} For each feature set, we look at the best-performing model (RF for aggregate and LR for BoW).

As the best performance for BoW features is achieved with Logistic Regression, we examine the $\beta$ coefficients 
of the model to find the words most useful in identifying each side of the vaccine debate, shown in Table \ref{tab:tfidffeatures}.
Each side is speaking critically, either about the other side, or about perceived problems.
The most indicative keywords for the vaccine skeptics refer to immigration (\textit{clandestino}, \textit{africano}), European politics (\textit{euro}, \textit{sovranita}, \textit{tedesco}, \textit{francese}), and left-wing Italian politics (\textit{pd}, \textit{sinistra}).
The keywords for vaccine advocates speak about the government (\textit{m5s}, \textit{lega}, \textit{ministro}, \textit{fascista}), and ``building a network'' (\textit{facciamorete} is an anti-government left-wing progressive hashtag).
Thus, we postulate that opinions about vaccination are often associated with particular political stances. 

\begin{table}[t]
\centering
\caption{BoW features having coefficients $\beta$ with highest magnitude on each side in Logistic Regression model.}
\label{tab:tfidffeatures}
\begin{tabular}{lr m{1em} lr}
\toprule
\multicolumn{2}{c}{\textbf{Advocates}} && \multicolumn{2}{c}{\textbf{Skeptics}} \\
\cmidrule(lr){1-2} \cmidrule(lr){4-5}
\textbf{Feature} & \textbf{$\beta$} && \textbf{Feature} & \textbf{$\beta$} \\
\midrule
facciamorete & 4.234 && clandestino & -4.265 \\
grillini & 3.836 && francese & -3.806 \\
ministro & 3.161 && sinistra & -3.803 \\
condonare & 3.072 && euro & -3.315 \\
oggi & 2.980 && sovranita & -3.269 \\
dire & 2.745 && piddini & -3.192 \\
persona & 2.510 && tedesco & -3.116 \\
leghista & 2.325 && africano & -3.040 \\
donna & 2.181 && banca & -2.967 \\
fascista & 2.109 && pd & -2.737 \\
\bottomrule
\end{tabular}
\end{table}

\begin{figure}[tbp]
\centering
\includegraphics[width=0.9\linewidth]{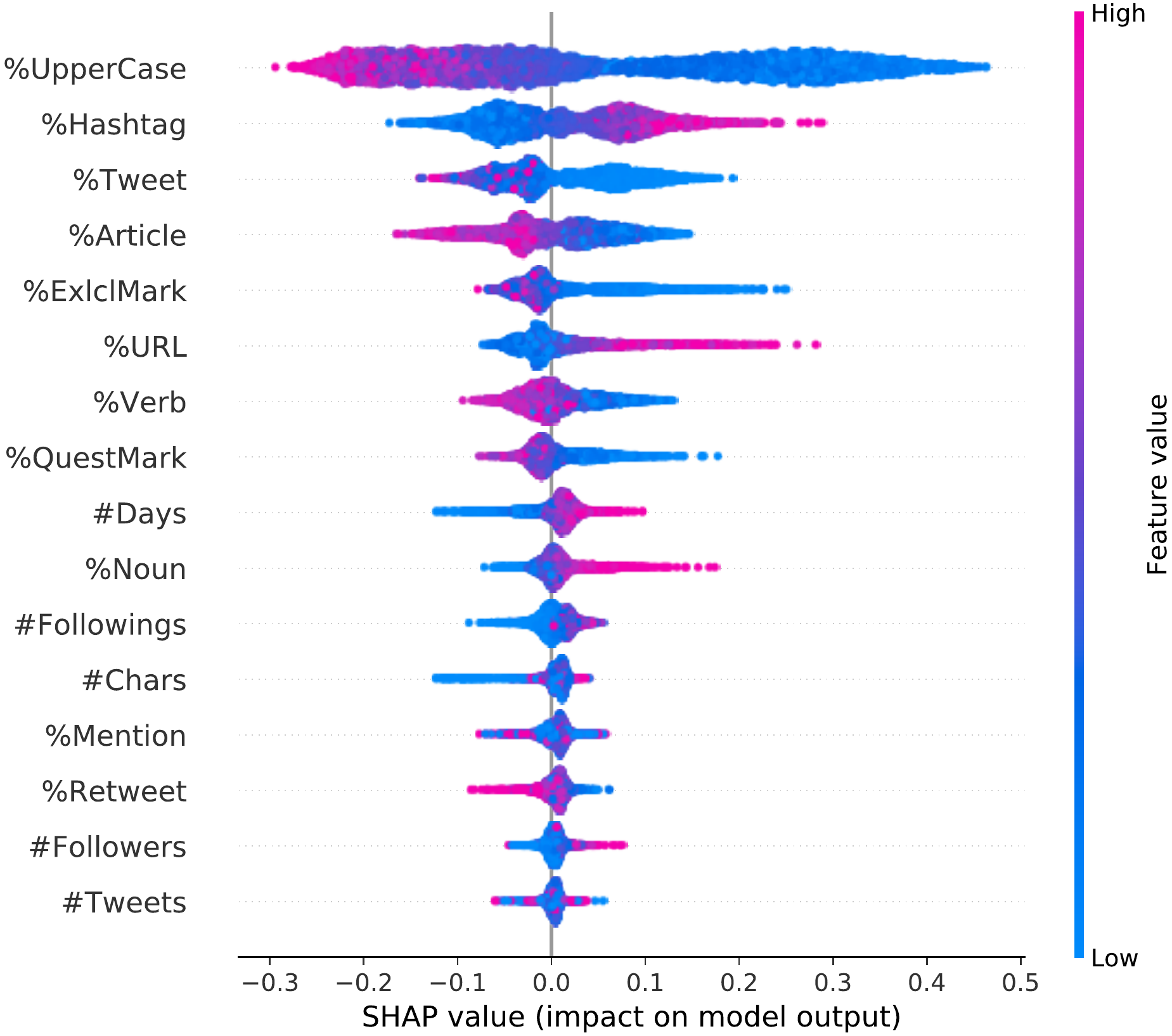}
\caption{Shapley additive explanations (SHAP) values for the aggregate features used in the random forest model. For each feature a distribution of SHAP values across the dataset is shown, with each point being a user (best seen in color).}
\label{fig:shap_plot}
\end{figure}

To understand the importance of the aggregate features, we use SHAP (Shapley additive explanations) values~\cite{lundberg2017unified}.
These values can be interpreted similarly to the $\beta$ of the Logistic Regression.
A summary plot of SHAP values for the entire dataset is shown in Figure~\ref{fig:shap_plot}.
Each point represents a user, thus for each feature the figure shows the distribution of SHAP values across the dataset.
Wider distribution indicate a larger absolute impact of the feature in the overall classification, while the color of each point encodes the feature value (low or high).
A feature with high values corresponding to positive SHAP values (to the right) is positively correlated with advocates (e.g., `\%Hashtag').
Conversely a feature with high values corresponding to negative SHAP values (to the left) is positively correlated with skeptics (e.g., `\%UpperCase').
The most predictive feature for skeptics is the use of uppercase letters, which represents a ``shouting'' rhetoric device typically associated with populist messages (such as pleas for ``maximum sharing'')~\cite{baldwin2019technological}.
We also find that skeptics use fewer hashtags\footnote{Section~``Debate Content \& Context'' refers to the main dataset, while this result derives from the historical data, hence the discrepancy. Possibly, the increased use of hashtags by skeptics in the main data is a response to the introduction of the vaccination law.}
 and URLs, tweet more, and use a higher proportion of articles, verbs, exclamation and question marks.
In addition, they tend to retweet more, and to have a younger account age.
Yet again, we discover cues of a picture where skeptics are a small but very vocal minority in the vaccination debate.

\textbf{Out-of-network Validation.} The prediction experiments in this section have looked at users who can be classified by using the retweet network partitioning, thus biasing our data to users who retweet others.
To examine the applicability of the data-driven prediction approach to users outside the GCC of the retweet network, we apply these models (Logistic regression trained on BoW features and Random Forest on aggregate features) to the users who can not be automatically labeled (those outside \users, the retweet network GCC).
We then sample $300$ users and have 5 Italian language speakers familiar with the topic domain label their leaning by looking at their vaccine-related tweets.

Unlike in the manual labeling of GCC users, this time a significant fraction of users does not display enough signal to manually detect their stance on vaccination, thus resulting in a 3-label test set: $64.7\%$ advocates, $23.0\%$ skeptics, and $12.3\%$ ``none''.
A 2-way Cohen's Kappa excluding ``none'' class produces a $\kappa$=0.628, whereas the 3-way one is $\kappa$=0.469, which indicates that users outside \users are more difficult to label.
Note the stronger imbalance in this sample (excluding ``none''), $73.6\%$ advocate and $26.4\%$ skeptic.
It is hard to judge from a small sample, but it could be the case that skeptical users are more likely to participate socially in the Twitter discussion.
By applying the models to the users with detectable stances, the aggregate features model performs at an accuracy of $0.776$ and the BoW model at $0.867$, a marked improvement over $0.736$ majority baseline, but a drop from the performance on users in \users.

When examining the agreement between the two classifiers, if we exclude users with manual label ``none'' , the agreement is $83.3\%$, but including those identified as ``none'' it drops to $78.4\%$.
This difference suggests the two algorithms disagree more on users labeled as ``none''.
This also indicates that the model does not overfit the data and in fact the prediction accuracy is low when there is not enough signal to provide a stable classifier, and that is despite having more historical data available (an average of \num{2473} historical tweets for users $\not\in \users$, compared to \num{878} tweets for users $\in \users$).
Observe that here we do not label for users being explicitly ``hesitant'', as the distinction between being vague about one's opinion and vocally undecided is a fine one.
Still, it is more likely that the more ``silent'' population could be more receptive to a public health information campaign.

\section{Discussion and Conclusions}
\label{sec:discussion}

In this work, we show that a structural network approach is a highly accurate methodology to identify the stance of users toward vaccines in Italian Twitter. 
In fact, we find cases in which network information proves to be vital in the detection of a user's opinion on the matter of vaccination, such as in the case of the former health minister, Giulia Grillo (@giuliagrillom5s), who is labelled as a skeptic.
While the content she has posted does not explicitly state an opinion, her tweets have been retweeted by 71 accounts we have identified as skeptic. 
In the past, Grillo has agreed with the notion that pharmaceutical companies use Italian population as lab subjects, advising vaccination instead only in ``case of need''\footnote{https://www.facebook.com/DavideFaraone/videos/ 1815267551861867/}.

Furthermore, our analysis allows us to identify the asymmetry in the interactions between and within the two communities.  
Not only do we show that the two sides of the debates tend to ignore each other's content, but the vaccine advocates do not even \emph{mention} the users from the skeptic side (see Topology of the echo chambers Section). 
This behavior may be due to vaccine advocates avoiding imbuing the skeptic side with legitimacy by engaging with them in an open debate, a tactic which however potentially leaves concerns voiced by the skeptics unanswered.
The divide is further seen in the follower network, thus implying that even outside the vaccination debate, the information sources the two sides consume are separate. 
This separation potentially supplies disjoint views of reality, different contextual frames, informational resources, and social norms.

These multiple dimensions of echo chambers within the Italian vaccination debate display a differentiation from earlier observations.
During the US political debate in 2010, \citet{conover2011political} found mention networks to be much less polarized than the retweet ones. 
However, when analyzing tweets from the US vaccine debate, the skeptics were found to be largely disconnected from the advocates~\cite{monsted2019algorithmic}. 
The extent of polarization was not quantified though.
The Dutch vaccination debate as seen through Twitter data presented a complex relationship between various players~\cite{lutkenhaus2019mapping}, thus identifying an ``anti-establishment'' group.
Unlike our observations for the case of Italy, the Dutch vaccine advocates actively engaged with the anti-establishment group, by providing evidence and responding to myths and misconceptions -- a strategy that the Italian counterparts may want to adopt (although such interventions may also backfire and reinforce the original belief \cite{horne2015countering}). 
An examination of cross-country communication and media influence in the global vaccination debate is an exciting future research direction.

Obviously, the conclusions of this work are limited to Italy and its unique political situation at the time of the data collection. 
Studies of other locales and cultures will need to both customize the language and keywords necessary for the data collection (differently from other work which conflates use of English word ``vaccines'' with worldwide interest in the topic~\cite{bello2017detecting}).
In addition, in this paper we simplify the diversity of stances on vaccination, which range from fine points on the medical side, such as concerns about vaccination scheduling and necessity of vaccination for rare diseases, to legislative control over the vaccination process, 
to conspiracy theories about purposeful secret sterilization campaigns. 
Any intervention will need to be adjusted to the particular political and social context of the objections to vaccination (in some countries politics may not even play a role), as well as to specific arguments against  vaccination (and possible misconceptions).

\textbf{Implications.} 
Given the recent push to ``vaccinate against hesitancy''~\cite{nature2019time} we propose several implications for future interventions, as informed by the findings of this study. 
First and foremost, we find a lack of engagement from the mainstream with the vaccination skeptic community, with mentions flowing mostly in the opposite direction.
Users peripheral to the central core of vaccine hesitancy clique may be especially susceptible to a targeted education campaign~\cite{horne2015countering}, and even if those in the core may not be easily swayed, the critique of their stances may be observed by the more ``silent'', less committed users. 

Second, political discussions and personal values such as the ``freedom of choice" dominate the vaccination debate, as the content analysis shows. 
These findings are also confirmed in the Italian vaccination debate as seen via Facebook data~\cite{Kalimeri2019}, where epistemic and ideological beliefs act as obstacles to the acceptance of scientific evidence. 
These viewpoints may be best tackled in these domains, instead of via purely scientific argumentation.

Third, our results suggest that URLs shared by skeptics are significantly less diverse than information cited by the pro-vaccination group, with a clear prevalence of YouTube videos followed by Facebook links. 
The YouTube platform was recently shown to provide ample anti-vaccination content in Italy~\cite{donzelli2018misinformation,covolo2017arguments}; both studies showed that even though most of the videos were positive in tone, those that disapproved of immunization were the most liked and shared.
Future interventions may want to utilize YouTube for messaging and engagement.

Overall, we contribute to a more profound understanding of how our digital ecosystem influences our access to vaccine-related information through selective exposure. 
The effect of opinion polarization is an increasingly important social phenomenon; 
we hope that our study will provide valuable insights for public health communication policymakers regarding potential outreach to those expressing concerns about vaccination on Twitter.

\section{Acknowledgments}
\label{sec:acknowledgment}
The authors  acknowledge support from the Lagrange Project of the Institute for Scientific Interchange Foundation (ISI Foundation) funded by Fondazione Cassa di Risparmio di Torino (Fondazione CRT).

\bibliographystyle{aaai}
\bibliography{vaccines}
\end{document}